\documentclass[12pt]{article}

\usepackage[margin=1in]{geometry}
\usepackage[version=4]{mhchem}
\usepackage[T1]{fontenc}
\usepackage{tabularx}
\usepackage{color}
\usepackage{soul}
\usepackage{authblk} 
\usepackage{setspace}
\usepackage{titlesec}
\usepackage{graphicx}
\usepackage[utf8]{inputenc} 

\newcommand{\myparallel}{{\mkern3mu\vphantom{\perp}\vrule depth 0pt\mkern2mu\vrule depth 0pt\mkern3mu}}

\title{Optical absorption in hexagonal-diamond Si and Ge nanowires: insights from STEM-EELS experiments and \textit{ab initio} theory}

\author[1]{Luiz H. G. Tizei}
\author[2]{Michele Re Fiorentin}
\author[3,4]{Thomas Dursap}
\author[3]{Theodorus M. van den Berg}
\author[1]{Marc Túnica}
\author[5]{Maurizia Palummo}
\author[1]{Mathieu Kociak}
\author[3]{Laetitia Vincent}
\author[1,*]{Michele Amato}

\affil[1]{Université Paris-Saclay, CNRS, Laboratoire de Physique des Solides, 91405 Orsay, France}
\affil[2]{Department of Applied Science And Technology (DISAT), Politecnico di Torino, C.so Duca degli Abruzzi 24, 10129, Torino, Italy}
\affil[3]{Université Paris-Saclay, CNRS, Centre de Nanosciences et de Nanotechnologies, F-91120, Palaiseau, France}
\affil[4]{Present address: IMEC, Kapeldreef 75, 3001 Leuven, Belgium}
\affil[5]{INFN, Dipartimento di Fisica, Università degli studi di Roma Tor Vergata, via della Ricerca Scientifica 1, Rome, 00133, Italy}
\affil[*]{Corresponding author: michele.amato@universite-paris-saclay.fr}

\date{} 

\begin{document}

\maketitle

\begin{abstract}
Hexagonal-diamond (2H) group IV nanowires are key for advancing group IV-based lasers, quantum electronics, and photonics. Understanding their dielectric response is crucial for performance optimization, but their optical absorption properties remain unexplored. We present the first comprehensive study of optical absorption in 2H-Si and 2H-Ge nanowires, combining high-resolution STEM, monochromated EELS, and \textit{ab initio} simulations. The nanowires, grown in situ in a TEM as nanobranches on GaAs stems, show excellent structural quality: single crystalline, strain-free, minimal defects, no substrate contamination, enabling access to intrinsic dielectric response. 2H-Si exhibits enhanced absorption in the visible range compared to cubic Si, with a marked onset above 2.5 eV. 2H-Ge shows absorption near 1 eV but no clear features at the direct bandgap, as predicted by \textit{ab initio} simulations. A peak around 2 eV in aloof-beam spectra is attributed to a thin 3C-Ge shell. These findings clarify structure–optical response relationships in 2H materials.
\end{abstract}

The recent advances in the fabrication techniques of hexagonal-diamond Si, Ge, and Si$_{1-x}$Ge$_{x}$ nanowires (NWs)~\cite{FadalyNATURE2020,HaugeNL17,HaugeNL2015}, (2H-Si, 2H-Ge, and 2H-Si$_{1-x}$Ge$_{x}$ NWs in the Ramsdell notation) and of their homojunctions~\cite{TangNANOSCALE2017,VincentNL2015,FasolatoNL2018,KaewmarayaJPC2017}, have opened new possibilities for enhancing the otherwise limited optical properties of silicon through selective crystal phase tuning~\cite{GalvaoEPJB2020,AmatoNL2016}.
In the last few years, the growth of Si and SiGe nanowires, showing the 2H phase, has been achieved in several different ways: by employing the crystal transfer method with various precursors\cite{FadalyNATURE2020,HaugeNL17,HaugeNL2015}, through strain-induced\cite{VincentNL2015} and photo-induced\cite{RodichkinaCEC2019} transformation processes, by using plasma-assisted vapor-liquid-solid (VLS) growth\cite{HeNANOSCALE2019,TangNANOSCALE2017} and molecular beam epitaxy (MBE)~\cite{DudkoCGD2021}. Furthermore, the 2H crystal phase control has been successfully proved also for 2H-Si and SiGe NW branches~\cite{LiNANOTECH2022,LamonNL2025}, 2H-Si nano-ribbons\cite{QiuSR15} and 2H-SiGe quantum wells~\cite{PeetersNC2024}.
The key interest in studying such nanostructures lies in the ability to modulate their intrinsic material properties, thereby enabling stronger and allowed optical transitions at the bandgap~\cite{PrioloNATNANO2014}. A comprehensive understanding of the properties of these novel optically active semiconductors could pave the way for the development of group IV-based lasers~\cite{vanTilburgJAP2023,vanTilburgCOMMPHYS2024}, as well as their potential applications in quantum electronics and photonics, including monolithically integrated electron-photon and spin-photon interfaces. Indeed, as predicted by \textit{ab initio} calculations~\cite{CartoixaNL2017}, recent photoluminescence (PL) measurements have shown that for Ge concentrations $x$ above 0.65, 2H-Si$_{1-x}$Ge$_{x}$ alloyed NWs are light-emitting semiconductors in the 1.5 $\mu$m -- 3.4 $\mu$m range~\cite{FadalyNATURE2020}. This functionality can also be extended below 1.5 $\mu$m by using 2H-SiGe quantum wells~\cite{PeetersNC2024}. 
These two seminal works have achieved the long-standing goal of fabricating high-quality group IV nanostructures suitable for optoelectronics, addressing a challenge that has eluded the community for decades~\cite{PrioloNATNANO2014}. Simultaneously, their findings have stimulated a rigorous discussion to understand the fundamental physical mechanisms governing the absorption and emission properties of 2H-Si-based materials. Several \textit{ab initio} studies suggested that increasing the Ge concentration $x$ in 2H-Si lowers the minimum of the
conduction band at $\Gamma$ until the bandgap becomes direct for
Ge concentrations larger than 0.65~\cite{CartoixaNL2017,BorlidoPRM2021}. Moreover, in agreement with PL measurements, the bandgap of these materials is associated with an allowed optical transition in the range 0.65 $<$ $x$ $<$ 1 and can be continuously tuned over this spectrum~\cite{BorlidoPRM2021,FadalyNATURE2020}. However, in contrast with experimental findings, more recent \textit{ab initio} simulations have demonstrated that pure 2H-Ge is a pseudodirect bandgap material with only a weakly dipole active transition between the highest conduction band and the lowest valence band at the $\Gamma$ point~\cite{RodlPRM2019}. 
This means that contrary to what was observed in experiments, the optical emission close to the pseudo-direct gap of 2H-Ge should be very low and not experimentally detectable. Interestingly, similar violations of the optical selection rules have been observed in wurtzite GaP~\cite{AssaliNL2013}, wurtzite GaAs~\cite{AhtapodovNL2012}, and wurtzite InP nanowires~\cite{PereraNL2013}. For 2H-Ge, a possible explanation of this discrepancy has been ascribed to crystal imperfections that could perturb the symmetry and induce a substantial modification of the selection rules compared to the perfect 2H crystal phase~\cite{vanACSPHOT2024,BroderickARXIV2024}. Therefore, since PL measurements are ambiguous and generally permit access only to the first excited state, absorption measurements are needed to characterize the full high-energy optical properties.

This letter focuses on a concerted experimental and theoretical effort to fully understand the absorption properties of individual 2H-Si and 2H-Ge NWs. 
We first fabricated single crystalline 2H-Si and 2H-Ge nanowires appearing as nanobranches on a wurtzite GaAs NW by the VLS growth mode. This nanobranch configuration appears to be beneficial compared to the core/shell configuration~\cite{FadalyNATURE2020} in avoiding the impact of the GaAs nanowire in the acquired spectra. Additionally, the nanowires were grown \textit{in situ} to tune in real time the growth parameters, and the as-grown samples were characterized successively in various (scanning) transmission electron microscopes ((S)TEM) for structural, chemical, and electron energy loss spectroscopy (EELS) analyses without any manipulation that may deteriorate the NWs. Due to the small density and the variations of morphology or structure between NWs or within individual NWs, a macroscopic absorption experiment was not appropriate to measure their dielectric properties. We, therefore, combined high spatial and spectral resolution of aberration-corrected STEM and monochromated EELS in the low-loss regime to investigate the optical excitation response of such nanostructures, identifying the main features in their absorption spectra. These measurements were then compared with first-principles calculations of the electronic and optical properties of 2H-Si and 2H-Ge to fully understand the mechanisms governing electronic excitations. 

Both theory and experiments show that the absorption onset lies below 2.7 eV for 2H-Si NWs. At higher energies, 2H-Si NWs show a dominant absorption peak at 3.7 eV, confirming the results of previous \textit{ab initio} calculations~\cite{RodlPRB2015}.
In contrast, the absorption of Ge is much broader between 2 and 4 eV. Interestingly, in the case of 2H-Ge NWs, the intersecting and aloof-beam EELS measurements give substantially different results. In particular, the absorption measured in the aloof-beam configuration presents a dominant peak around 2 eV. Thanks to the comparison with our parameter-free optical spectra simulations, we could ascribe the origin of this peak to the presence of a thin 3C-Ge shell covering the 2H-Ge core. As reported in a previous study~\cite{LiNANOTECH2022}, this is due to the radial vapor-solid (VS) growth that can occur on $\{0001\}$ sidewalls. For the intersecting EELS configuration, no absorption signature can be seen below 2 eV and close to the direct bandgap of the 2H-Ge phase at around 0.3 eV. This suggests that, as proposed in several theoretical works~\cite{RodlPRM2019,BorlidoPRM2023,BroderickARXIV2024}, the observed emission in 2H-Ge may be linked to point, or extended defects, nanostructuring, strain, or surfaces/interfaces, which could potentially influence the dipole transitions at the direct bandgap~\cite{vanACSPHOT2024}. It is important to note that the emission coefficient in unstrained 2H-Ge is very low~\cite{BroderickARXIV2024}, but any perturbation, such as strain or defects, could enhance it. However, this remains speculative, and further accurate experimental investigation is required to confirm these hypotheses.

This coupled experimental and theoretical study represents the first comprehensive report on optical absorption in pure 2H-Si and 2H-Ge nanowires, making a significant contribution to advancing the understanding of electronic excitations in this emerging class of semiconductors. The growth and characterization of the hexagonal phases were achieved through a combination of unique materials, ultra-high-resolution EELS experimental techniques, and advanced atomistic modeling. This approach provides valuable insights into the field and advances it by offering a novel framework for understanding the dielectric response in hexagonal-diamond group IV nanowires. NWs were grown \textit{in situ} in modified TEM for MBE growth on special MEMS. These MEMS were then characterized in another TEM,  with < 10 meV EELS spectral resolution. NWs of both cubic and hexagonal phases are present in the samples. Therefore, only the capability of analyzing the structure of individual NWs, coupled with high spectral resolution EELS, allowed for the measurement of the optical response of the hexagonal phases.

\begin{figure}[htbp]
\centering
\includegraphics[width=0.4\textwidth]{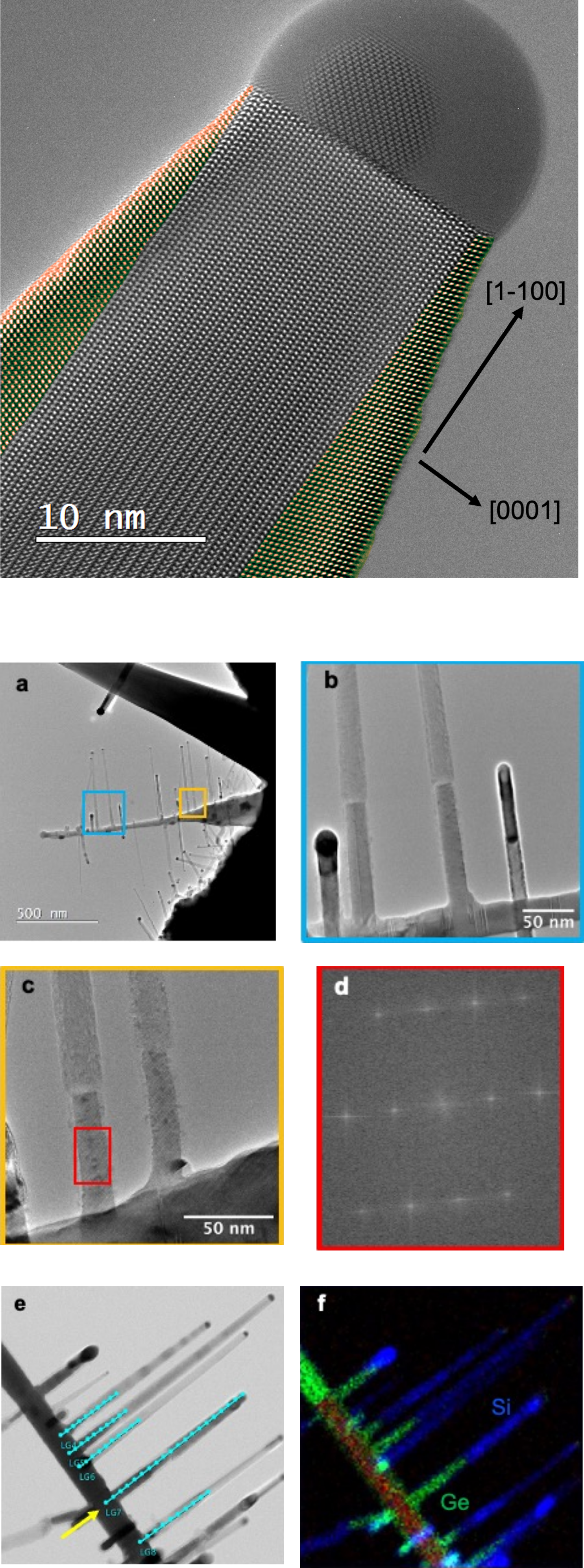}
\caption{Top panel: HRTEM micrograph of a Ge nanowire captured during \textit{in situ} growth. The crystal structure below the Au catalyst droplet is perfectly hexagonal in epitaxy with the GaAs wurtzite nanowire trunk. The cubic structure on the $\{0001\}$ sidewalls is due to a vapor solid overgrowth using digermane. Bottom panel: (a) TEM overview of GaAs trunk with nanobranches on the sidewalls. (b) and (c) HRTEM of the selected area (blue and orange boxes). (d) FFT of the nanobranch in (c). (e) HAADF-STEM micrograph of Ge-Si nanobranches. (f) STEM-EDS elementary mapping.}
\label{figure_growth}
\end{figure}

We synthesized high-quality 2H-Si and 2H-Ge NWs using vapor-liquid-solid growth on wurtzite GaAs stems. The growth was carried out \textit{in situ} in a TEM, allowing real-time control of the nanowire morphology~\cite{LiNANOTECH2022} (see Supporting Information (SI)). 
2H-Ge NWs grown along the $\{1\bar{1}00\}$ direction presented a hexagonal structure that is perfectly replicated below the Au catalyst (see top panel of Fig.~\ref{figure_growth}). A vapor-solid overgrowth resulting in a cubic shell (highlighted in orange in Fig.~\ref{figure_growth}) expanding during growth can be observed on the $\{0001\}$ sidewalls.
The growth of Si nanowires as nanobranches on GaAs NWs was even more challenging due to an enhanced spreading of Au droplets on the sidewalls under SiH$_{4}$ (see SI). The straight and pure Ge/Si heterostructures obtained with this approach are shown in Fig.~\ref{figure_growth}(a). The contrast between the Ge and Si parts is visible on the micrograph shown in Fig.~\ref{figure_growth}(b) and Fig.~\ref{figure_growth}(c). The Ge stem nanobranch is about 100 nm, while the Si nanobranch is about 350 nm. We note that using SiH$_{4}$ to grow the Si nanowires induces a very thin overgrowth on the sidewall, compared to the Ge nanowire shown in the top panel of Fig.~\ref{figure_growth}. The grown nanowires were further analyzed using STEM. Energy dispersive X-ray spectroscopy (EDS) was performed using a JEOL (2200~FS operating at 200~keV), showing the elementary composition and confirming the pure Si and Ge segments (see Figs.~\ref{figure_growth}e and~\ref{figure_growth}f). We want to emphasize that our samples are inherently hexagonal diamond crystals. A previous study~\cite{LiNANOTECH2022} first demonstrated that Si and Ge nanowires synthesized using this method exhibit the 2H phase with no stress, negligible extended and point defects, and no substrate contamination. This exceptional crystal quality makes these nanowires particularly well-suited for fundamental studies to measure and understand their optical response.

As shown in several theoretical works~\cite{RodlPRB2015,RodlPRM2019,KaewmarayaJPC2017,AmatoNL2019}, both 2H-Si and Ge crystals are single-element hexagonal-diamond structures belonging to the P6$_{3}$/\textit{mmc} space group and presenting two distinct dielectric polarization axis: one related to the in-plane directions (\textit{a} and \textit{b} axis) and the other along the out-of-plane direction (\textit{c} axis). This crystal anisotropy induces particular features in the electronic band structures (see SI). 
2H-Si is an indirect semiconductor whose fundamental bandgap of 0.92 eV is along the $\Gamma$-$M$ line. Another relevant minimum is present at the $K$ point while the direct bandgap, 3.4 eV in 3C-Si, is here reduced to 1.75 eV (see left panel of Fig.~S1 of the SI). On the other hand, 2H-Ge has a direct bandgap of 0.26 eV at $\Gamma$ with another relevant minimum at $M$ (see right panel of Fig.~S1 of the SI).

EELS is known to be related to the optical properties of nanostructures. In the case of a hypothetical isotropic material of dielectric constant $\epsilon(\omega)$, the EELS response for a beam intersecting the material is proportional to $Im(-1/\epsilon(\omega))$. Far from any bulk plasmon excitations, this, in turn, is essentially proportional to $Im(\epsilon(\omega))$. In other words, far from plasmonic excitation, the bulk EELS response reveals optical transitions. In the case of an anisotropic material, the expressions are mathematically more cumbersome, but essentially, the conclusion stays identical (see, for example, Refs.~\cite{Bolton1995,Yeh2021}). In the case of multilayered materials, the weight contribution of each material will be proportional to their thickness along the electron beam path. When the beam is not intersecting, the exact EELS response will depend on the geometry of the nanostructure and the beam position. However, in the case of excitations far from plasmons, this response will again be proportional to $Im(\epsilon(\omega))$ \cite{Boudarham2012}. Multilayer materials will also show optical responses from the different materials. The case of anisotropic material has not been thoroughly covered, but recent work \cite{Yeh2021} has shown that the EELS response is again sensitive to optical transitions. Therefore, for a better interpretation of our experimental results, we calculated the macroscopic dielectric functions and the corresponding loss function for both 2H-Si and 2H-Ge nanowires from first principles.

\begin{figure}[htbp]
\centering
\includegraphics[width=0.8\textwidth]{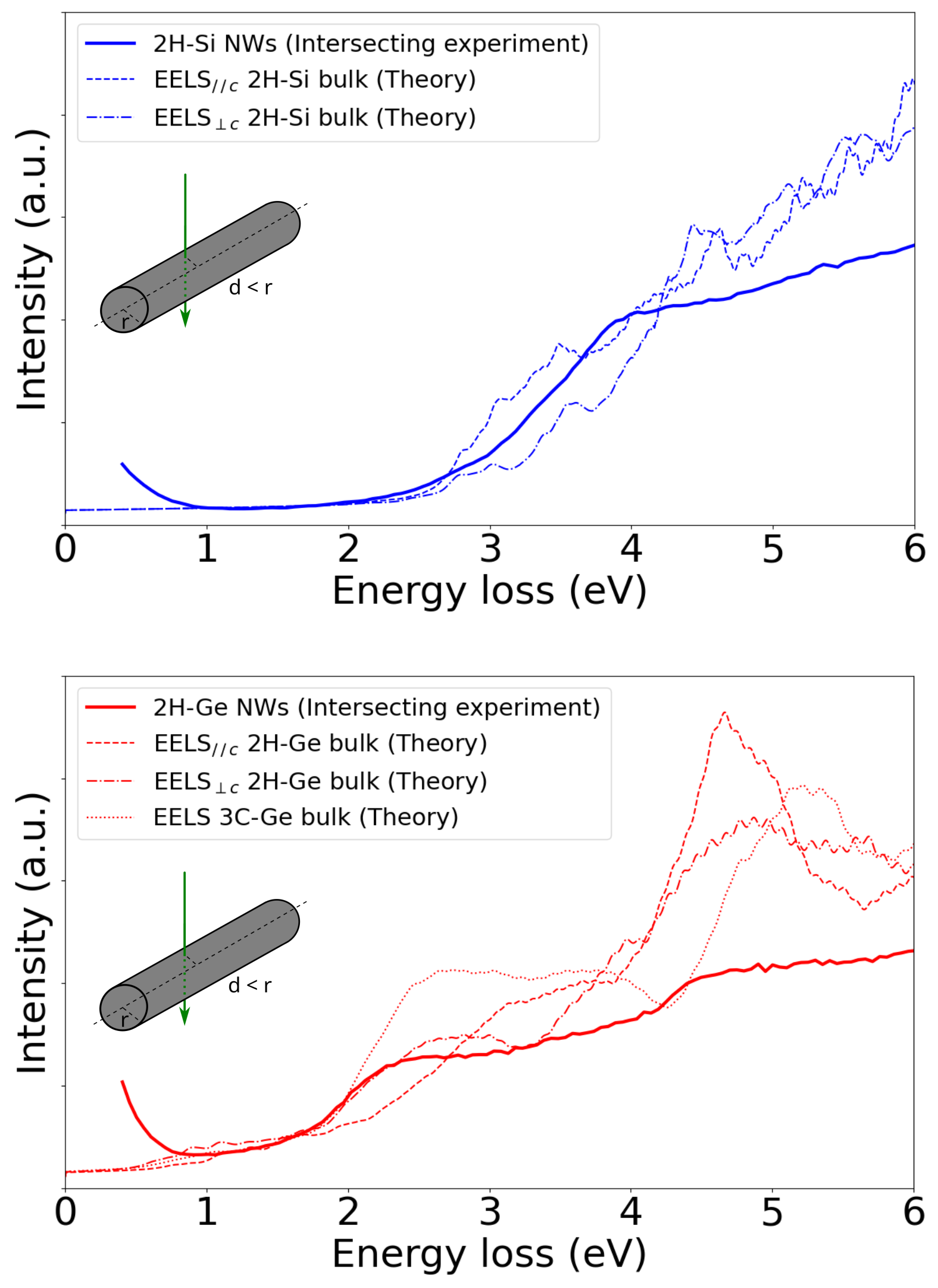}
\caption{Intersecting-specimen beam EELS spectra for 2H-Si NWs (blue continuous line in the top panel) and 2H-Ge NWs (red continuous line in the bottom panel). In the same plots, the loss functions calculated within many-body perturbation theory for in-plane and out-of-plane polarizations are reported for bulk 2H-Si (blue dashed and dash-dotted lines in the top panel) and bulk 2H-Ge (red dashed and dash-dotted lines in the bottom panel). The calculated loss function of bulk 3C-Ge is also reported for comparison with the other cases (red dotted line in the bottom panel). The inset in each panel shows a sketch of the intersecting-specimen beam EELS configuration mode.}
\label{bulk_eels}
\end{figure}

\begin{figure}[htbp]
\centering
\includegraphics[width=\textwidth]{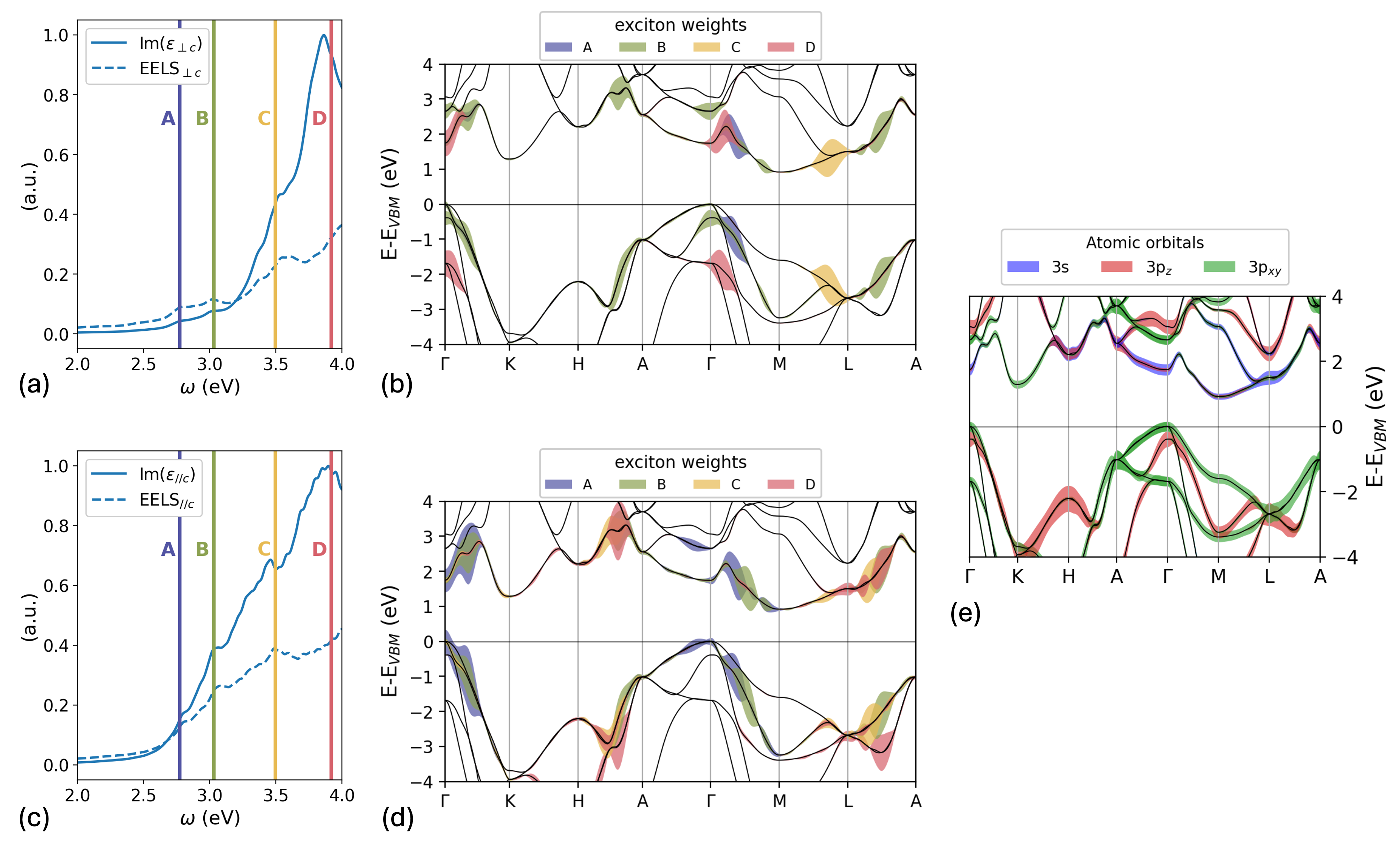}
\caption{\textbf{2H-Si}. Panels (a) and (c): dielectric and loss functions, perpendicular and parallel to $c$, respectively. Selected peaks in the spectra are marked by vertical lines and labeled A to D. Panels (b) and (d): weights of electronic transitions contributing to the corresponding peaks in panels (a) and (c), respectively. Panel (e): band structure of 2H-Si, with its $k$-resolved projections onto 3$s$, 3$p_{z}$ and 3$p_{xy}$ atomic orbitals highlighted with different colors.}
\label{label:2H-Si_exciton}
\end{figure}

\textbf{Intersecting-specimen EELS configuration}. A first comparison between measurements and theoretical results has been reported in Fig.~\ref{bulk_eels}, where the intersecting-specimen EELS spectra and the calculated loss function for Si and Ge NWs are shown. 
In the case of 2H-Si NWs (top panel of Fig.~\ref{bulk_eels}), the experimentally measured loss function (solid line) lies between the calculated in-plane and out-of-plane ones (dot-dashed and dashed lines). This could be ascribed to the large convergent angle between the beam and the sample, which allows the beam to excite transition along both polarizations. In this configuration, the absorption of 2H-Si starts at 2.77~eV and continuously starts to increase up to nearly 4 eV, as confirmed by the calculated spectra. At the same time, the spectral intensities remain very weak at low energies, close to the fundamental bandgap (whose theoretical value is 0.92~eV). In Figs.~\ref{label:2H-Si_exciton}(a) and (c), we label the most prominent features of the theoretical loss function and absorption spectra computed for in-plane and out-of-plane light polarization, respectively. The main electronic transitions contributing to the corresponding exciton states are highlighted in panels (b) and (c) with same-color shading. By analyzing peaks A and B in Fig.~\ref{label:2H-Si_exciton}(c), we can attribute the absorption increase above 2.5~eV to electronic transitions excited by out-of-plane polarized light, as shown in Fig.~\ref{label:2H-Si_exciton}(d). These transitions occur between the next-to-topmost valence band (VB-1) and the conduction band (CB) along the $\Gamma\rightarrow K$ and the $\Gamma \rightarrow M$ directions.
Looking at the $k$-resolved density of states projected onto Si atomic orbitals, in Fig.~\ref{label:2H-Si_exciton}(e), it is clear that the character of these transitions essentially derives from  3$p_{z}$~\textrightarrow~3$s$ and 3$p_{xy}$~\textrightarrow~3$s$ excitations. For better clarity, in Table~\ref{Table:dipole_orientations}, we report the energy and the orbital character of the analyzed electronic transitions. A comparison of the dielectric functions for the two light polarizations, Fig.~\ref{label:2H-Si_exciton}(a) and (c), points out the strong anisotropy of the system.    

\begin{figure}[htbp]
\centering
\includegraphics[width=\textwidth]{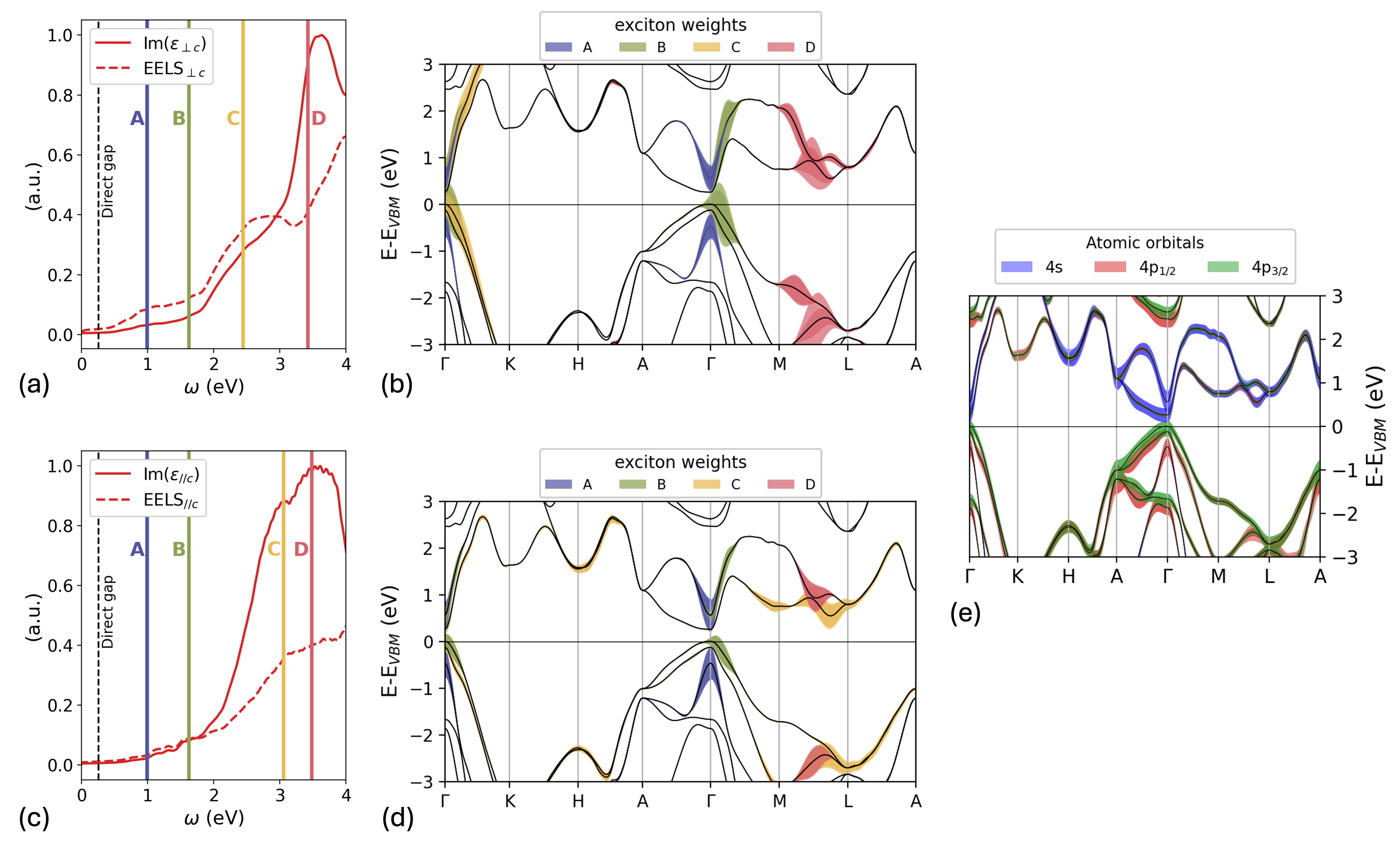}
\caption{\textbf{2H-Ge}. Panels (a) and (c): dielectric and EEL functions, perpendicular and parallel to $c$, respectively. Selected peaks in the spectra are marked by vertical lines and labeled A to D. Panels (b) and (d): weights of electronic transitions contributing to the corresponding peaks in panels (a) and (c), respectively. Panel (e): Band structure of 2H-Ge accounting for SOC, with its $k$-resolved projections onto 4$s$, 4$p_{1/2}$ and 4$p_{3/2}$ atomic orbitals highlighted with different colors.}
\label{label:2H-Ge_exciton}
\end{figure}

As for 2H-Ge (see bottom panel of Fig.~\ref{bulk_eels}), the intersecting-specimen EELS measurement reveals that the absorption starts to increase around 1 eV in agreement with our calculated theoretical spectra (dashed lines in the bottom panel), and the results of previous calculations (see for instance Ref.~\cite{BorlidoPRM2023}). In Fig.~\ref{label:2H-Ge_exciton}(a) and (c), we mark the main features of the theoretical absorption and loss function of 2H-Ge. Most importantly, no noticeable structures appear in the theoretical spectra at energies close to the direct bandgap of 0.26~eV, marked by the vertical dashed lines in panels (a) and (c). This is because, for both polarizations, the transition between VB and CB at $\Gamma$ is dipole forbidden, as the conduction band minimum in 2H-Ge derives from the zone folding of the $L$ point in the cubic-diamond of Ge. 
As shown in Fig.~\ref{label:2H-Ge_exciton}(b) and (d), the intensity increase in the spectra corresponding to state A are related to electronic transitions at $\Gamma$ between VB-2 and CB+1 for both polarizations\cite{RodlPRM2021}. Panels (b) and (d) further confirm that the character of the transitions from 1 eV to 3 eV essentially derives from the transitions between the three highest valence bands and the two lowest conduction bands. Finally, the $k$-resolved density of states, reported in panel (e), shows that, in the same energy range, the corresponding transitions are essentially composed by 4$p_{3/2}$~\textrightarrow~4$s$ excitations and can have both an in-plane and out-of-plane nature. The energy, polarization, and orbital character of the analyzed transitions are reported in Table~\ref{Table:dipole_orientations}.

\begin{figure}[htbp]
\centering
\includegraphics[width=0.8\columnwidth]{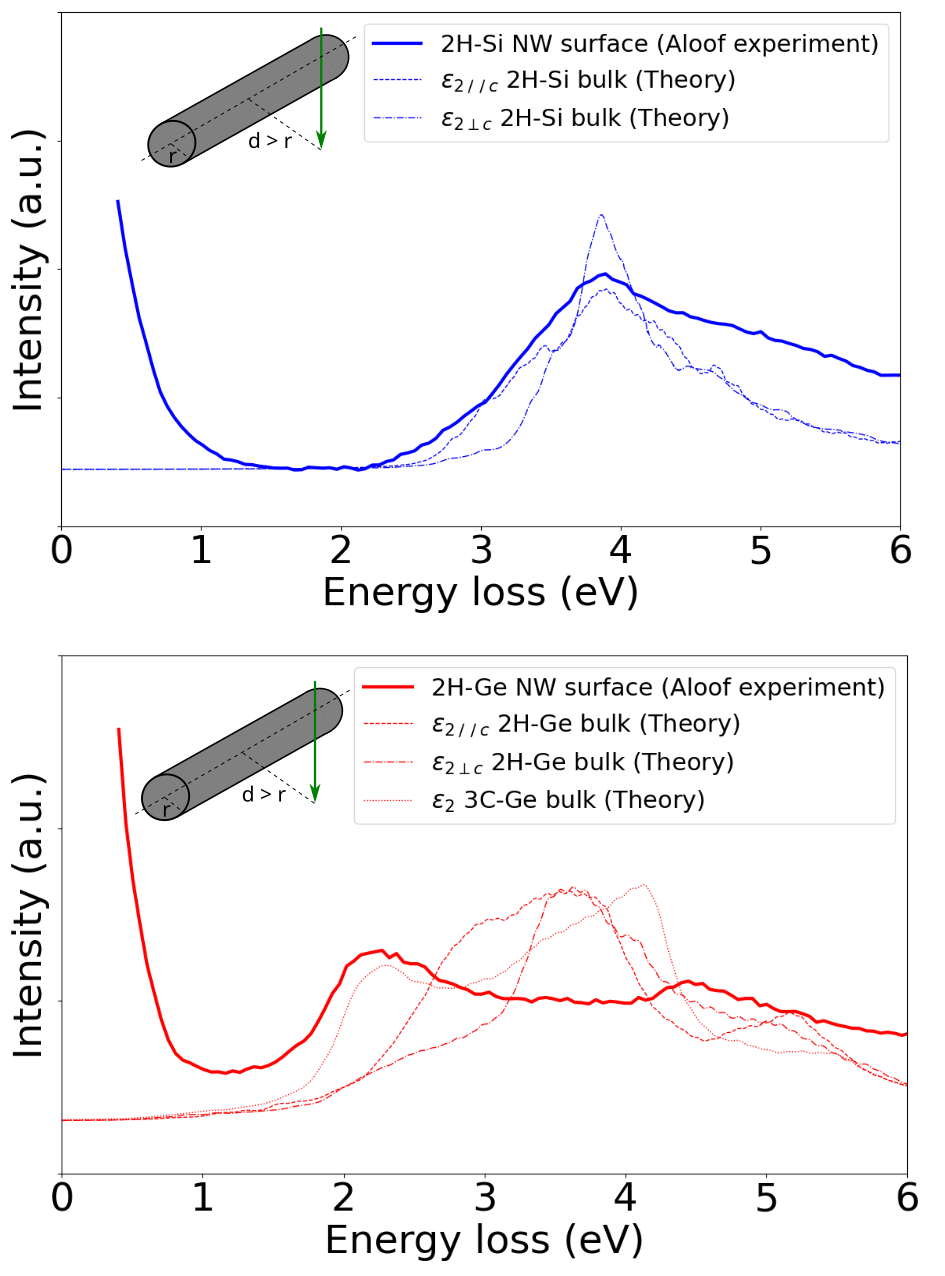}
\caption{Aloof-beam EELS spectra for 2H-Si NWs (blue continuous line in the top panel) and 2H-Ge NWs (red continuous line in the bottom panel). In the same plots, the imaginary part of the dielectric function, Im($\epsilon_{M}$), calculated within many-body perturbation theory, for in-plane and out-of-plane polarizations are reported for bulk 2H-Si (blue dashed and dash-dotted lines in the top panel) and bulk 2H-Ge (red dashed and dash-dotted lines in the top panel). The calculated loss function of bulk 3C-Ge is also reported for comparison with the other cases (bottom panel). The inset in each panel shows a sketch of the aloof-beam EELS configuration mode.}
\label{label:surface_eels_eps}
\end{figure}

\textbf{Aloof-beam EELS configuration}. The aloof-beam measurements are reported in Fig.~\ref{label:surface_eels_eps} for 2H-Si NWs and 2H-Ge NWs. Such an EELS configuration does not intersect the specimen. It is essentially a near-field spectroscopy that is very sensitive to the surface morphology of the sample, in particular in the case of nanostructures~\cite{CohenPRL1998}. The results for 2H-Si NWs (top panel of Fig.\ref{label:surface_eels_eps}) show an excellent agreement between the measured spectrum and the calculated dielectric functions. As in the case of the intersecting-specimen case, the measured spectrum lies between the in-plane and the out-of-plane components theoretically calculated (dashed lines in the top panel of Fig.~\ref{label:surface_eels_eps}). The measurement shows that absorption of 2H-Si NWs starts to increase for energies larger than 2 eV with a very high intense peak around 4 eV.
On the other hand, understanding the measured absorption for 2H-Ge NWs is more puzzling. Here, we observe a very intense peak at 2 eV (continuous line in the bottom panel of Fig.~\ref{label:surface_eels_eps}) in contrast to what is predicted by our \textit{ab initio} calculations (dashed lines in the bottom panel of Fig.~\ref{label:surface_eels_eps}). Interestingly, this first peak perfectly matches the first absorption peak calculated for bulk 3C-Ge, whose band structure, including SOC, is shown in the right panel of Fig.~S1 of the SI. Indeed, as reported in the previous sections and shown in Fig.~\ref{figure_growth}, the grown samples can present a thin 3C-Ge shell due to the competition between the axial VLS and the radial VS process growths. Since in the aloof configuration, the beam does not penetrate the specimen, the dielectric response at low energies is mainly dominated by the 3C-Ge shell covering the 2H-Ge core whose thickness is around 1.5~nm~\cite{LiNANOTECH2022}. This peak was absent in the case of the intersecting beam configuration because the bulk loss function was to be probed.

Thus, the experimental results and theoretical analysis demonstrate that Si nanowires exhibit a well-defined absorption peak at 3.7 eV. Experimentally, Ge nanowires show no absorption signature below 2 eV, near the direct bandgap. More specifically, around 0.3 eV, we estimate that no energy onset is detectable above the noise level (0.1\% of the 2 eV onset). As a result, our EELS experiments do not reveal substantial absorption in the far infrared. This is consistent with our theoretical calculations (see Fig.~S2 in the SI), which indicate that the oscillator strengths associated with the direct bandgap transition, though nonzero, are minimal. Interestingly, this contrasts with photoluminescence (PL) experiments, which have demonstrated strong far-infrared emission from 2H-Ge~\cite{FadalyNATURE2020}. The discrepancy between absorption and emission results could be due to extrinsic factors (such as stress or defects) influencing the PL emission~\cite{vanACSPHOT2024,RodlPRM2019,BroderickARXIV2024}. However, given the different experimental conditions \textemdash absorption being detected via EELS in individual, thin nanowires, and emission being detected in PL from macroscopic ensembles or larger nanowires \textemdash a direct comparison of these results on the same samples remains elusive.

We have grown and optically characterized the dielectric response of single crystal 2H-Si and 2H-Ge NWs through intersecting and aloof-beam STEM-EELS configurations. Measured optical absorption spectra have been successfully compared with \textit{ab initio} theoretical predictions calculated within many-body perturbation theory, including excitonic, non-collinear, and local-field effects. We found that 2H-Si NWs, due to their reduced direct bandgap (1.75 eV), have more significant absorption in the visible spectral range with respect to cubic-diamond Si. However, their absorption becomes significant only for energies larger than 2.5 eV. This experimental result confirms all the previous theoretical calculations~\cite{AmatoNL2016,RodlPRB2015,BorlidoPRM2023,BorlidoPRM2021}, which predicted a large absorption for 2H-Si NWs in the visible region, thus suggesting their potential for photovoltaic applications, in particular, when heterostructured with 3C-Si NWs~\cite{KaewmarayaJPC2017,RodlPRB2015,AmatoNL2019,AmatoNL2016}. The measurements performed on 2H-Ge NWs are strongly dependent on the beam-to-specimen orientation due to the complex morphology of the grown Ge samples. In particular, we found that the absorption increases around 0.9 eV in the intersecting-specimen EELS measurement, which agrees with the \textit{ab initio} theory. Interestingly, no absorption signature can be recognized for energies close to the direct bandgap. This result is of primary interest as the light emission from 2H-Ge and has been the object of intense debate in the last few years~\cite{FadalyNATURE2020,RodlPRM2019,vanACSPHOT2024,PeetersNC2024}. We both theoretically and experimentally confirm that the direct bandgap transition in 2H-Ge is extremely weak, which contradicts the previously observed light emission from 2H-Ge samples. It is important to note that our samples are high-quality, with minimal defects and no substrate contamination. This discrepancy has broader implications, particularly for systems where violations of optical selection rules have been observed, such as wurtzite GaP~\cite{AssaliNL2013}, wurtzite GaAs~\cite{AhtapodovNL2012}, and wurtzite InP nanowires~\cite{PereraNL2013}. The aloof-beam measurements show instead a very intense peak at 2 eV in contrast to what is predicted by our \textit{ab initio} simulations for 2H-Ge. By comparing this result with the theoretical absorption spectra of 3C-Ge, we ascribed the discrepancy to the presence of a thin 3C-Ge shell that can be produced during the VLS process. 

Besides providing theoretical insights, our study explores the fundamental absorption mechanisms in hexagonal-diamond Si and Ge nanowires. Given the growing interest in high-Ge-content hexagonal alloys for developing group IV-based lasers~\cite{vanTilburgCOMMPHYS2024,vanTilburgJAP2023}, as well as their potential applications in quantum electronics and photonics, these findings represent an important initial step toward understanding and engineering the dielectric response of this emerging class of semiconductors. Moreover, a deeper understanding of optical absorption in 2H-Ge could offer valuable insights into the electronic excitations of other pseudo-direct bandgap nanostructures~\cite{AhtapodovNL2012,PereraNL2013,AssaliNL2013}.


\section*{Supporting Information}
Materials growth, TEM characterization, electronic band structure calculations, optical absorption calculations, EELS measurements. 

\section*{Acknowledgement}
All the authors acknowledge the ANR HEXSIGE project (ANR-17-CE30-0014). M.A. and M. T. acknowledge the ANR AMPHORE project (ANR-21-CE09-0007) and the ANR TULIP project (ANR-24-CE09-5076). L. V. acknowledges funding from the European Union’s Horizon 2020 research and innovation program under grant agreement No. 964191 (OptoSilicon) and No. 101080022 (ONCHIPS), the ANR for funding the NANOMAX ETEM and the CHROMATEM microscope through the TEMPOS grant (10-EQPX-0050).
L.V. and T.D. thank Labex Nanosaclay (ANR-10-LABX-0035) for their financial support. We also acknowledge the CIMEX at École Polytechnique (Palaiseau, France) for hosting the NANOMAX microscope. Part of the high-performance computing resources for this project were granted by the Institut du développement et des ressources en informatique scientifique (IDRIS) under the allocations AD010914974 and AD010915077 via GENCI (Grand Equipement
National de Calcul Intensif).
M.R.F. and M.P. acknowledge the Italian Research Center on High Performance Computing, Big Data and Quantum Computing (ICSC), funded by the European Union - NextGenerationEU and established under the National Recovery and Resilience Plan (PNRR), as well as high-performance computing resources provided by CINECA 
through the ISCRA initiative. M.P. also acknowledges INFN Time2quest funding.

\bibliographystyle{unsrt}
\bibliography{manuscript}
\pagebreak

\newpage

\newpage

\newpage
\begin{table}[htbp]
\caption{Calculated optical transitions below 4 eV, from valence to conduction states, and their polarization orientation with respect to the $c$ axis in 2H-Si and 2H-Ge.}
\centering
\begin{tabular}{cccc}
\hline\hline
Crystal & Energy (eV) & Orientation & Assignment \\
\hline\hline
\textbf{2H-Si} & 2.77 & $\perp$ $\&$ $\myparallel$ & 3$p_{z}$~\textrightarrow~3$s$ \\
              & 3.02 & $\perp$ $\&$ $\myparallel$ & 3$p_{z}$~\textrightarrow~3$s$ \\
              & 3.49 & $\myparallel$ & 3$p_{xy}$~\textrightarrow~3$p_{xy}$ \\
              & 3.56 & $\perp$ & 3$p_{xy}$~\textrightarrow~3$s$ \\
              & 3.82 & $\perp$ & 3$p_{xy}$~\textrightarrow~3$s$ \\
              & 3.92 & $\myparallel$ & 3$p_{xy}$~\textrightarrow~3$s$ \\
\textbf{2H-Ge} & 0.99 & $\perp$ $\&$ $\myparallel$ & 4$p_{1/2}$~\textrightarrow~4$s$ \\
              & 1.62 & $\perp$ $\&$ $\myparallel$ & 4$p_{3/2}$~\textrightarrow~4$s$ \\
              & 2.44 & $\perp$ & 4$p_{3/2}$~\textrightarrow~4$s$ \\
              & 3.05 & $\myparallel$ & 4$p_{3/2}$~\textrightarrow~4$s$,~4$p_{3/2}$ \\
              & 3.43 & $\perp$ & 4$p_{3/2}$~\textrightarrow~4$s$,~4$p_{3/2}$ \\
              & 3.48 & $\myparallel$ & 4$p_{1/2}$~\textrightarrow~4$s$,~4$p_{3/2}$ \\
\hline\hline
\end{tabular}
\label{Table:dipole_orientations}
\end{table}

\newpage

\newpage

\newpage

\newpage
\begin{figure}[htbp]
\centering
\includegraphics[width=\textwidth]{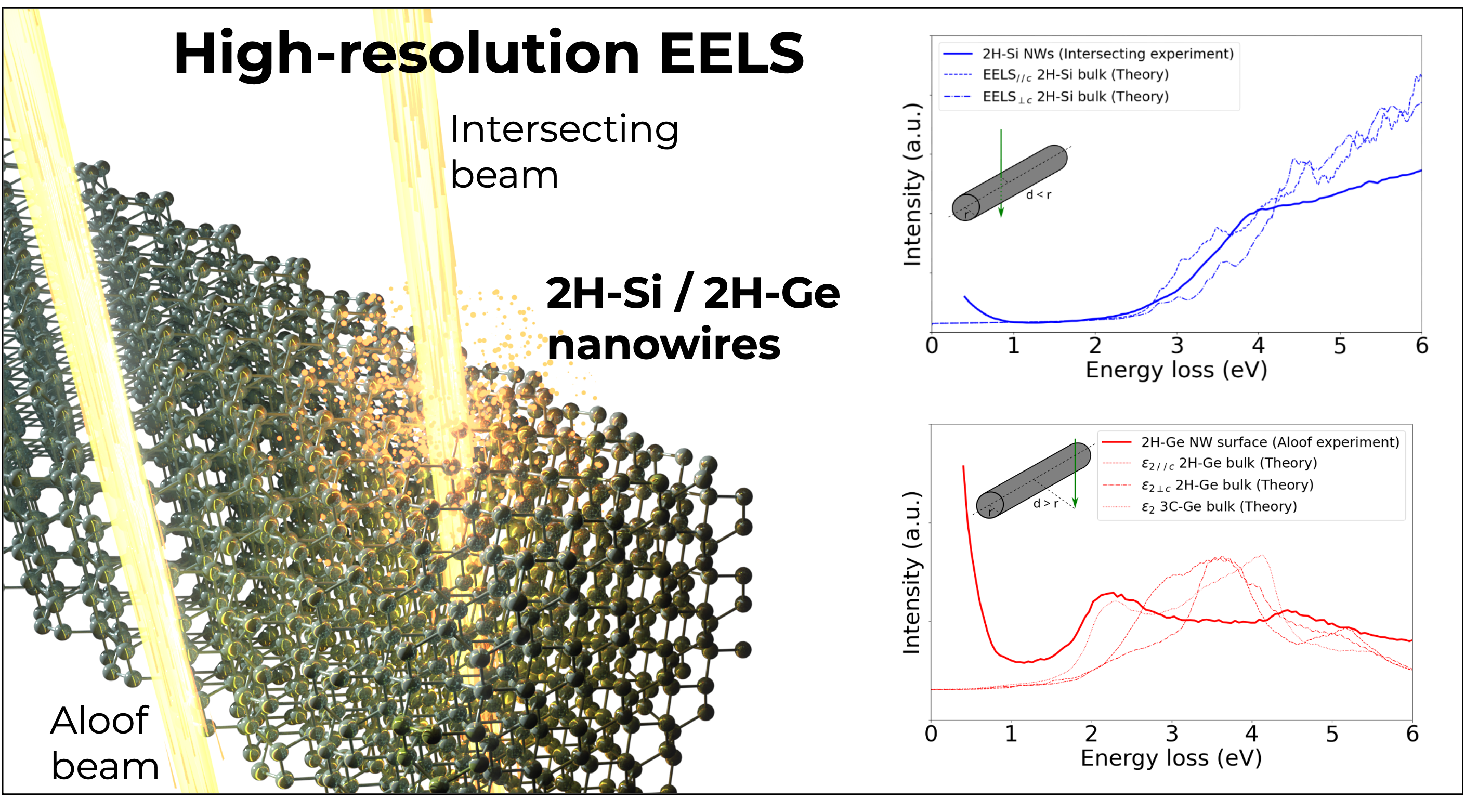}
\caption*{TOC graphic.}
\end{figure}

\end{document}